\documentclass[aps,prx,amsmath,amssymb,reprint,superscriptaddress]{revtex4-1}
\usepackage[latin1]{inputenc}
\usepackage{graphicx}
\usepackage{amsmath}
\usepackage{dcolumn}
\usepackage{bm}
\usepackage{bbold}
\usepackage{color}
\usepackage{amsfonts}
\usepackage{amssymb}
\usepackage{mathrsfs}
\usepackage{tabularx}
\usepackage{braket}
\usepackage{mathtools}
\usepackage{soul}			

\usepackage{caption}
\usepackage{subcaption}

\begin{document}

\title{Temperature coefficients of crystalline-quartz elastic constants over the cryogenic range [4 {\rm K}, 15 {\rm K}]}

\author{J\'{e}r\'{e}my Bon}
\email{jeremy.bon@femto-st.fr}
\affiliation{FEMTO-ST Institute, Univ. Bourgogne franche-Comt\'{e}, CNRS, ENSMM, 26 rue de l'\'{E}pitaphe, 25030 Besan\c{c}on, France}

\author{Serge Galliou}
\email{serge.galliou@femto-st.fr}
\affiliation{FEMTO-ST Institute, Univ. Bourgogne franche-Comt\'{e}, CNRS, ENSMM, 26 rue de l'\'{E}pitaphe, 25030 Besan\c{c}on, France}

\author{Roger Bourquin}
\email{roger.bourquin@ens2m.fr}
\affiliation{FEMTO-ST Institute, Univ. Bourgogne franche-Comt\'{e}, CNRS, ENSMM, 26 rue de l'\'{E}pitaphe, 25030 Besan\c{c}on, France}


\date{\today}

\begin{abstract}

This paper brings out results of a measurement campaign aiming to determine the temperature coefficients of synthetic quartz elastic constants at liquid helium temperature. The method is based on the relationship between the resonance frequencies of a quartz acoustic cavity and the elastic constants of the material. The temperature coefficients of the elastic constants are extracted from experimental frequency-temperature data collected from a set of resonators of various cut angles, because of the anisotropy of quartz, measured on the very useful cryogenic range [4 {\rm K}-15 {\rm K}]. The knowledge of these temperature coefficients would allow to further design either quartz temperature sensors or conversely frequency-temperature compensated quartz cuts. With extremely low losses, lower than $10^{-9}$ for the best ones, key applications of such devices are ultra-low loss mechanical systems used in many research areas including frequency control and fundamental measurements. The Eulerian formalism is used in this study to identify the temperature coefficients.

\end{abstract}

\maketitle

\section{Introduction}

Bulk acoustic wave (BAW) piezoelectric devices are widely used in research and industry to make temperature sensors, acoustic cavities or filters for several applications. Recently it has been demonstrated that quartz acoustic cavities can exhibit very high quality factors, up to a few billions for the best ones, at liquid helium temperature~\cite{galliouScRip2013}. In these conditions they become very attractive for various experiments in fundamental physics~\cite{Muller2016, Goryachev:2014aa}, as well as for hybrid quantum systems~\cite{Goryachev:2013aa, AspelmeyerNature2010}, optomechanics~\cite{Aspelmeyer2014}, etc. This interest is amplified by the fact that commercial cryorefrigerators are now easy to use. In the field of ultra stable frequency standards, the advent of superconducting devices also offers new interesting opportunities. The coupling of those cavities exhibiting extremely low mechanical losses with SQUID (Superconducting QUantum Interference Device) is a first step towards cryogenic ultra-stable frequency sources~\cite{GoryachevSQUID2014}. Nevertheless, progress in this field suffer of the lack of knowledge regarding the physical constants of the acoustic cavity material, the crystalline quartz, at cryogenic temperatures. The very first topic of interest is the behavior of elastic constants versus temperature. Although a set of temperature coefficients of the elastic constants of quartz already exists since 1975~\cite{Smagin}, it covers the very large temperature range $4 {\rm K}-400 {\rm K}$, and it turns out that it is not reliable within the very useful range $4 {\rm K}-15 {\rm K}$ commonly achieved by standard cryocoolers. In addition, these coefficients have been determined from the fast and slow (quasi-)shear propagation modes usually denoted B and C respectively, and not from the (quasi-)longitudinal mode, denoted A, the one which exhibits the highest quality factors at low temperatures.

Several methods can be used to determine the thermal coefficients. It can be from phase velocity measurements by pulse echo overlap methods, resonant ultrasound spectroscopy or combined resonance techniques, or by means of laser ultrasonic techniques. However, application of these methods at low temperature arises technical problems. Alternatively, one way consists in probing the temperature sensitivity of a resonance frequency from a dedicated acoustic device. Indeed, the acoustic resonance frequency of a mechanical system is a function of its elastic coefficients which are themselves temperature dependent. This method is particularly relevant for a quartz crystal device: the temperature sensitivity of the elastic coefficients of the quartz material can be determined by identifying the frequency-to-temperature coefficients of a plano-convex quartz resonator, i.e. a trapped energy resonator, whose frequency of overtone order $n$ derives from the basic relationship $f_n=\frac{n}{2h}\sqrt{\frac{C_{e}}{\rho}}$ where $2h$ is a thickness, $C_{e}$ an elastic coefficient as described below, and $\rho$ the quartz density. This latter approach is used here to extract the temperature coefficients of quartz stiffness within $4 {\rm K}-15 {\rm K}$ from various quartz-cut with different orientations relative to the crystallographic reference.

\section{Method}

The Eulerian formalism can be used to determine the thermal coefficients of the elastic coefficients $C_{ijkl}$ because the linear thermal expansion coefficients of quartz are known at the operating temperatures~\cite{Barron}.

Strictly speaking, according to the Eulerian description, the problem should be solved in the so-called final state~\cite{Dulmet2001},~\cite{Dulmet2001bis} by adding the mechanical strains caused by the acoustic vibration to the temperature strains induced by the temperature change from room temperature to cryogenic temperature. However, as an available approximation, the problem can be solved in the intermediate state by taking into account the strains induced by the very small vibration-caused displacement magnitude, typically in the order of a nanometer.

In the Euler formalism the strain-stress relationship is given by the Cauchy tensor expressed, for the intermediate state, by $T_{ij}=C_{ijkl}\frac{\partial u_k}{\partial x_l}$ where $C_{ijkl}$ is the elastic coefficient tensor and $u_k$ a component of the displacement vector. The eigenfrequencies of the acoustical modes of the mechanical structure obviously depend on the elastic coefficients $C_{ijkl}$ that are temperature dependent.
To bring up the thermal coefficients, the elastic coefficients can be written in the form of a Taylor series:
\begin{equation}
\begin{array}{l}
C_{ijkl}(T)=C_{ijkl}(T_{ref})\left[1+\sum_k T^{(k)} C_{ijkl} \right. \vspace{8pt}\\
\hspace{48pt} \left.(T-T_{ref})^k\right]
\end{array}
\label{eq:elastic_coef}
\end{equation}

where $C_{ijkl}(T_{ref})$ are the elastic coefficients at the reference temperature $T_{ref}$, and $T^{(k)} C_{ijkl}$ the order $k$ thermal coefficients defined by:

\begin{equation}
T^{(k)} C_{ijkl} = \frac{1}{k!}\frac{1}{C_{ijkl}(T_{ref})}\frac{d^{k}C_{ijkl}}{dT^{k}}
\label{eq:Taylor_coef}
\end{equation}

\begin{figure}[t!]
	\centering
		\includegraphics[width=3.25in]{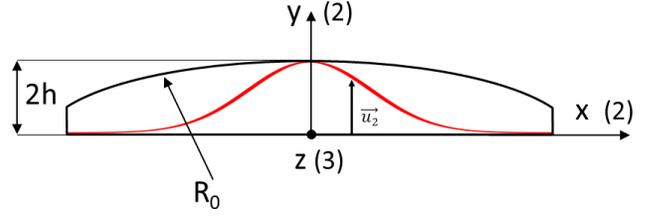}
	\captionof{figure}{Plano-convex resonator. In such a trapped-energy device the vibration displacement $u_2$ along the Y-axis has a gaussian distribution \cite{Tiersten1986}.}
	\label{fig:plano_convex_scheme}
\end{figure}

\begin{figure}[t!]
	\centering
		\includegraphics[width=3.25in]{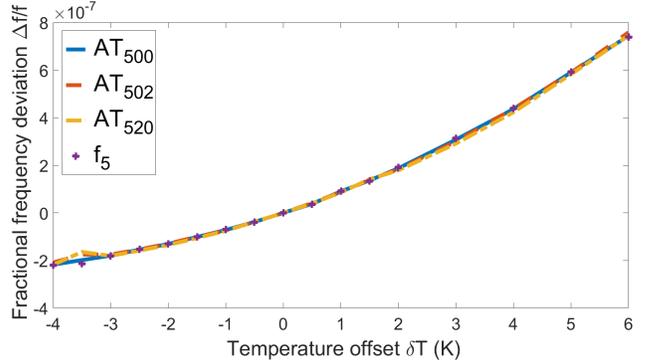}
	\captionof{figure}{Fractional frequency deviation of $f_5$ $f_{502}$ and $f_{520}$ versus the temperature offset with respect to $T_{ref}=8 \rm{K}$, in the case of the single-rotated Y-cut called AT-cut, as an example. Plots of the $5^{th}$ order resonance frequency $f_5$ of the infinite flat plate calculated from the plano-convex resonator frequencies $f_{5mp}$ are in very good agreement with \eqref{eq:Plate_convex_eigenfrequency}.}
	\label{fig:eigenfrequencies_AT-cut_plano_convex}
\end{figure}

The eigenfrequencies of the plano-convex resonators measured in this study (see Fig.~\ref{fig:plano_convex_scheme}) can be expressed according to the Stevens-Tiersten model~\cite{Tiersten1986}:

\begin{equation}
    \begin{array}{l}
f_{nmp}^2=f_n^2\left[1+\frac{1}{n\pi}\sqrt{\frac{2h}{R_0}}\left(\sqrt{\frac{M_n}{C_e}}(2m+1)\right.\right. \vspace{8pt}\\
\hspace{32pt} \left.\left.+\sqrt{\frac{P_n}{C_e}}(2p+1)\right)\right]
    \end{array}
\label{eq:Euler_eigenfrequency}
\end{equation}

where index $n$ is the overtone number and $m,p$ the number of modal lines in the resonator plane, $M_n$ and $P_n$ the dispersion constants, $2h$ the resonator thickness, $R_0$ the curvature radius of the convex face, and $f_n$ the eigenfrequencies of the corresponding infinite flat plate. As shown in Fig.~\ref{fig:eigenfrequencies_AT-cut_plano_convex}, the latter can usually be computed accurately from the eigenfrequencies of the plano-convex resonator by combining both anharmonic modes $n02$ and $n20$, and the corresponding quasi-harmonic one $n00$ as follows:
\begin{equation}
		f_n^2=\frac{(6f_{n00}^2-f_{n02}^2-f_{n20}^2)}{4}
\label{eq:Plate_convex_eigenfrequency}
\end{equation}

As a consequence, the temperature coefficients $T^{(k)}C_{ijkl}$ of the elastic constants can just be determined from the temperature coefficients of the eigenfrequencies of the corresponding infinite flat plate. These eigenfrequencies are given by \cite{Tiersten1986} :
\begin{equation}
{f_n^\xi}^2(T)=\frac{n^2}{4h^2(T)}\frac{\overline{C_{e}^\xi}(T)}{\rho (T)}\left(1-\frac{8k_\alpha^2(T)}{n^2 \pi^2}-r \right)
\label{eq:Plate_eigenfrequency}
\end{equation}

where $\overline{C_{e}^\xi}$ is the effective elastic coefficient, $k_\alpha^2$ the coupling electromechanical coefficient, $h$ the resonator thickness (about 1 mm; the electrode thickness is neglected because it is small enough, a few hundreds of nanometers, to neglect the wave propagation inside the electrodes), $\rho$ the mass density of quartz, $n$ the overtone range and $r$ the ratio between masses of the resonator and electrodes (typically $10^{-3}$). $\xi$ denotes A, B or C, referring to the extensional propagation mode, the high-speed thickness mode, and the low-speed thickness mode respectively.

Actually, terms inside the brackets of~\eqref{eq:Plate_eigenfrequency} can be reduced to unity in the $T^{(k)}C_{ijkl}$ calculation. Indeed, the third term $r$ is a constant independent of temperature, and then is not involved in this calculation. Regarding the second term inside the brackets of~\eqref{eq:Plate_eigenfrequency}, values of the electromechanical coefficients $k_\alpha^2$ are known only at room temperature, but are close to $10^{-4}$. In addition, as any physical constant, their thermal sensitivity are smaller at low temperature than at room temperature. Thus, this second term can also be reasonably neglected especially for overtone orders greater than 3.

The effective elastic coefficients $\overline{C_{e}^\xi}$ are eigenvalues in the Christoffel theory, given by \cite{Auld} in the case of a piezoelectric material:
\begin{equation}
\overline{\Gamma_{il}}=C_{ijkl}^E n_j n_k + \frac{(e_{pij} n_p )(e_{qkl} n_q )}{(\epsilon_{jk}^S n_j n_k )}
\label{eq:Christoffel_1}
\end{equation}

where $C_{ijkl}^E$ are the elastic coefficients measured at constant electric field, $e_{pij}$ and $e_{qkl}$ the piezoelectric constants, $\epsilon_{jk}^S$ the dielectric constant at constant strain and $n_\beta$ (with $\beta=p,q,j$ or $k$) the directions of propagation. The second term of~\eqref{eq:Christoffel_1} is linked to the electromechanical coefficient, and it turns out that it can also be neglected in comparison with the first one.

To sum up, it results from the simplifications described above that piezoelectricy is not involved anymore. According to these hypotheses the issue is then limited to its pure mechanical nature, and it turns out that: $\overline{C_{e}^\xi}\approx C_{e}^\xi$ and $\overline{\Gamma_{il}} \approx \Gamma_{il}$.

Coefficients $C_{e}^{\xi}$ are eigenvalues of the Christoffel tensor\cite{Royer}:
\begin{equation}
\Gamma_{il} {}^\circ u_l = \rho V^2 {}^\circ u_i
\end{equation}
with $\Gamma_{il} = C_{ijkl} n_j n_k$, and where $V$ and ${}^\circ u_l$ are respectively the phase velocity and polarization of the acoustic wave.
Terms of the symmetrical tensor $\Gamma_{il}$ are defined in the crystalline class 32 of quartz as:
\begin{equation}
    \begin{array}{lll}
			\Gamma_{il} = C_{i11l} n_1^2+C_{i22l} n_2^2+C_{i33l} n_3^2+(C_{i12l}+ \vspace{4pt} \\
			\hspace{25pt} C_{i21l})n_1 n_2+ (C_{i13l}+C_{i31l}) n_1 n_3+(C_{i23l}+ \vspace{4pt} \\
			\hspace{25pt} C_{i32l}) n_2 n_3
    \end{array}
\end{equation}
where the index $i$ and $l$ varies from 1 to 3.

The cubic shape of the frequency-temperature characteristic observed experimentally implies that the thermal coefficients of the elastic constants should be provided up to the $3^{rd}$ order. These thermal sensitivities can be obtained by differentiating equation~\eqref{eq:Plate_eigenfrequency} with respect to temperature:

\begin{equation}
    \begin{array}{lll}
        T^{(1)}C_{e} \approx 2T^{(1)}f+T^{(1)}\rho+2T^{(1)}h \vspace{6pt} \\
				T^{(2)}C_{e} \approx \frac{1}{2}[T^{(1)}C_{e}]^2-[T^{(1)}f]^2-\frac{1}{2}[T^{(1)}\rho]^2- \vspace{6pt} \\
				\hspace{42pt} [T^{(1)}h]^2+2T^{(2)}f+T^{(2)}\rho+2T^{(2)}h \vspace{6pt} \\
				T^{(3)}C_{e} \approx 2T^{(3)}f-2T^{(1)}fT^{(2)}f+\frac{2}{3}T^{(1)}f+ \vspace{6pt} \\
				\hspace{42pt} \frac{1}{3}[T^{(1)}\rho]^3-\frac{1}{3}[T^{(1)}C_{e}]^3+\frac{2}{3}[T^{(1)}h]^3- \vspace{6pt} \\
				\hspace{42pt} T^{(1)}\rho T^{(2)}\rho-2T^{(1)}hT^{(2)}h+2T^{(3)}h+ \vspace{6pt} \\
				\hspace{42pt} T^{(1)}C_{e}T^{(2)}C_{e}+T^{(3)}\rho
    \end{array}
		\label{eq:Thermal_coeff}
\end{equation}

where $T^{(k)}\gamma=\frac{1}{\gamma !}\frac{d^{k}\gamma}{dT^{k}}$, with $\gamma = C_e, h, \rho$, or $f$.

The temperature coefficients of the thickness $2h$ are simply the expansion coefficients of the quartz material within $4.5 K-15 K$, with $T_{ref}=8 K$.  Their numerical values have been computed by a third order fitting of data provided by~\cite{Barron} (see Tab.~\ref{tab:table4}), and according to the proper resonator cut angles. 

\begin{table} [h!]
\caption{\label{tab:table4}Interpolated values of expansion coefficients from ~\cite{Barron} for the quartz material over the cryogenic range [4.5K-15K]. $\alpha_{\perp}$ and $\alpha_{//}$ are the linear expansion coefficients along the X-axis and Z-axis respectively.}
\begin{ruledtabular}
\begin{tabular}{cccc}
 & $\alpha^{(1)}, 10^{-8}$ & $\alpha^{(2)}, 10^{-8}$ & $\alpha^{(3)}, 10^{-8}$ \\
\hline
$\alpha_{\perp}$ $(\alpha_{11})$ & $0.9651$ & $0.5160$ & $0.1416$ \\
$\alpha_{//}$ $(\alpha_{33})$ & $-0.5134$ & $-0.1202$ & $0.0241$ \\
\end{tabular}
\end{ruledtabular}
\end{table}

The temperature coefficients of the mass density $\rho$ also depend on the previous expansion coefficients, and can be expressed as follows~\cite{Dulmet}:

\begin{equation}
    \begin{array}{lll}
        T^{(1)}\rho = -(2\alpha_{11}^{(1)}+\alpha_{33}^{(1)}) \vspace{6pt} \\
				T^{(2)}\rho = -(2\alpha_{11}^{(2)}+\alpha_{33}^{(2)}-3(\alpha_{11}^{(1)})^2-\vspace{6pt} \\
				\hspace{44pt} (\alpha_{33}^{(1)})^2-2\alpha_{11}^{(1)}\alpha_{33}^{(1)}) \vspace{6pt} \\
				T^{(3)}\rho = -(2\alpha_{11}^{(3)}+\alpha_{33}^{(3)}-2\alpha_{11}^{(1)}(3\alpha_{11}^{(2)}+\vspace{6pt} \\
				\hspace{44pt} \alpha_{33}^{(2)})-2\alpha_{33}^{(1)}(\alpha_{11}^{(2)}+\alpha_{33}^{(2)})+ \vspace{6pt} \\
				\hspace{44pt} 3(\alpha_{11}^{(1)})^2 \alpha_{33}^{(1)}+2\alpha_{11}^{(1)}(\alpha_{33}^{(1)})^2+ \vspace{6pt} \\ 
				\hspace{44pt} 4(\alpha_{11}^{(1)})^3+(\alpha_{33}^{(1)})^3)
    \end{array}
\end{equation}

Regarding the frequency differentiation, it is obtained from a $3^{rd}$ order fitting of a set of experimental frequency-temperature plots recorded from  quartz resonators of various cut angles described in Tab.~\ref{tab:table3}. In the Eulerian description, six frequency-temperature graphs are necessary to determine the three temperature coefficients of each of the seven elastic coefficients of the tensor $C_{ijkl}$ written in matrix notation as:

\begin{equation}
\begin{pmatrix}
\centering
C_{11} & C_{12} & C_{13} & C_{14} & 0 & 0 \\
C_{12} & C_{11} & C_{13} & -C_{14} & 0 & 0 \\
C_{13} & C_{13} & C_{33} & 0 & 0 & 0 \\
C_{14} & -C_{14} & 0 & C_{44} & 0 & 0 \\
0 & 0 & 0 & 0 & C_{44} & C_{14} \\
0 & 0 & 0 & 0 & C_{14} & C_{66} \\
\end{pmatrix}
\label{eq:matrix_Cijkl}
\end{equation}

In~\eqref{eq:matrix_Cijkl}, the elastic coefficients are given according to the specific rules of index contraction:

\begin{table} [h!]
\begin{tabular}{p{2.5cm}p{2.5cm}c}
\centering $(11) \leftrightarrow 1$ & \centering $(22) \leftrightarrow 2$ & $(33) \leftrightarrow 3$ \vspace{6pt} \\
\centering $(23)=(32) \leftrightarrow 4$ & \centering $(13)=(31) \leftrightarrow 5$ & $(12)=(21) \leftrightarrow 6$ \\
\end{tabular}
\end{table}

Cuts and resonance modes used for the calculation of the temperature coefficients are summarized in Tab.~\ref{tab:table1}. 

The temperature coefficients $C_e^{\xi}$ can be linked to the temperature behavior of the elastic coefficients by:

\begin{equation}
T^{(k)}C_e^{\xi}=\left[A_{\xi j}\right]^{-1} T^{(k)}C_{j}  
\end{equation}
Where $\left[A_{\xi j} \right] = \frac{C_{e}^{\xi}}{C_{j}}\frac{\partial C_{e}^{\xi}}{\partial C_{j}}$ and $C_j$ are the elastic coefficients $C_{ij}$ arranged in a column vector where index $j$ defines the component location in this vector. Values of the elastic coefficients $C_{ij}$ of $\left[A_{ij}\right]$ are those provided by \cite{Bechmann}. Although these values are given at room temperature, they can also be used at cryogenic temperature with a relative accuracy better than $10^{-3}$, the fractional frequency shifts observed from room temperature to cryogenic temperature operations being typically lower than $10^{-3}$.

\begin{table} [h]
\caption{\label{tab:table3}Elastic coefficients and quartz-cuts used for their respective determination. the X, Y, and Z-cuts are cuts whose planes are normal to the X, Y, and Z cristalline axis respectively. The AT-cut is a single rotated Y-cut, i.e. a cut whose cut plane, originally normal to the Y-axis, is rotated around the X-axis. The SC-cut is a doubly-rotated cut, i.e. a AT-cut rotated around $Z'$ the new Z-axis after the first rotation. The elastic coefficient $C_{66}$ is related to $C_{11}$ and $C_{12}$ as $C_{66}=(C_{11}-C_{12})/2$ \cite{Royer}.}
\begin{ruledtabular}
\begin{tabular}{ccc}
$\textit{Elastic coefficient}$ & $\textit{Quartz cut}$ & $\textit{Comments}$\\
\hline
$C_{11}$ & X & \\
$C_{12}$ & X & Calculated from $C_{11}$ and $C_{66}$ \\
$C_{13}$ & SC & \\
$C_{14}$ & AT & \\
$C_{33}$ & SC & \\
$C_{44}$ & Z & \\
$C_{66}$ & Y & \\
\end{tabular}
\end{ruledtabular}
\end{table}

\begin{table} [h!]
\caption{\label{tab:table1}Cuts of crystal plates and modes used to measure the corresponding resonance-frequency coefficients from a $3^{rd}$ order polynomial fit, with $T_{ref}=8$K. A, B and C modes are the (quasi-)longitudinal, fast (quasi-)shear and slow (quasi-)shear thickness modes respectively.}
\begin{ruledtabular}
\begin{tabular}{c p{1.8cm} ccc}
$\textit{Quartz cut}$ & \centering{$\textit{Mode}$} & $T^{(1)}f, 10^{-7}$ & $T^{(2)}f, 10^{-8}$ & $T^{(3)}f, 10^{-9}$ \\
\hline
AT & \centering{$C_{300}$, $C_{500}$} & 0.769 & 0.748 & 0.105 \\
SC & \centering{$A_{300}$, $A_{500}$} & -1.178 & -2.189 & -2.474 \\
SC & \centering{$B_{300}$, $B_{302}$, $B_{320}$, $B_{500}$} & -0.879 & -1.672 & -1.832 \\
X & \centering{$A_{100}$, $A_{1300}$, $A_{1500}$} & -0.143 & -0.451 & -0.581 \\
Y & \centering{$C_{300}$, $C_{500}$} & 1.339 & 1.869 & 1.289 \\
Z & \centering{$C_{300}$, $C_{500}$} & -1.465 & -2.662 & -3.571 \\
\end{tabular}
\end{ruledtabular}
\end{table}

\section{Experiment}

The temperature coefficients of the elastic tensor of the quartz material at liquid helium temperature have been calculated from the set of experimental frequency-temperature characteristics of various cut angles. Actually the resulting frequency coefficients in Tab.~\ref{tab:table1}, used to determine these temperature coefficients, are averaged values from measurements made with several overtone numbers of each tested cut.
 
\begin{figure}[t!]
	\centering
		\includegraphics[width=3.25in]{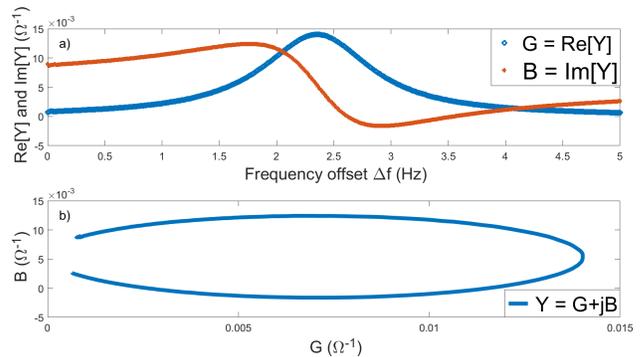}
	\captionof{figure}{\rm{a)} G-B plot from the network analyzer for the $1^{st}$ OT of a X-cut at 4.5K. $\Delta f$ is with respect to $f_{ref} = 4.580957 \rm{Hz}$. \rm{b)} Admittance circle from the experimental data above.}
	\label{fig:adm_circle_plot}
\end{figure}

The general overview of the measurement procedure can be summed up by the following steps:
$1/$ Resonators are realized in a set of plates of various cut angles  according to the conventional process: plate orientation with respect to the quartz crystal axis, plate cutting, rounding, lapping and polishing to obtain a 13 mm diameter plano-convex device, gold-electrode coating for piezo-electrically exciting/probing the acoustic cavity. The thickness of these resonators is such that the $3^{rd}$ overtone of the C-mode is close to 10 MHz at room temperature. 
Cuts used to take into account the quartz anisotropy comply with the IEEE-std 176-1978 \cite{IEEE}: AT ($0^\circ, -35.3^\circ$), SC ($22^\circ, -34^\circ$), X ($30^\circ, 0^\circ$), Y ($0^\circ, 0^\circ$), Z ($0^\circ, 90^\circ$). $2/$ To be tested, the resonators are embedded inside an oxygen-free cooper block screwed on the cold stage of a pulse-tube cryorefrigerator (a Sumitomo cryocooler SRP082B).  This block is cooled down in the vicinity of 4 K and temperature controlled within $\pm 3$ mK around the set temperature value when the steady state is reached. $3/$ For a given temperature, the resonator admittance $Y$ is measured around the interesting resonance frequencies, and over a frequency span properly chosen, by means of a network analyzer. However a calibration procedure should be followed before recording, to cancel the influence of the coaxial cable feeding the resonator inside the vacuum chamber from the network analyzer: this calibration involves three identical coaxial cables respectively ended by a standard 50 ohms resistor, an open circuit and a short circuit. The network analyzer (HP4195A) is locked to a hydrogen maser to get the frequency-stability performances of the latter. The analyzer provides a $0.001$ dB and $0.01^\circ$ resolution in magnitude and phase respectively, and its frequency resolution reaches $1$ mHz. In addition, amplitude-to-frequency nonlinear effects are avoided by limiting the excitation power. $4/$ The post-processing consists in plotting the admittance circle $Y=G+jB$ from the experimental data, magnitude and phase of $Y$, to get the resonance frequency with the best accuracy. Fig.~\ref{fig:adm_circle_plot} shows an example of such a G-B plot.



%

\section{Results}

\begin{figure}[t!]
	\centering
		\includegraphics[width=3.25in]{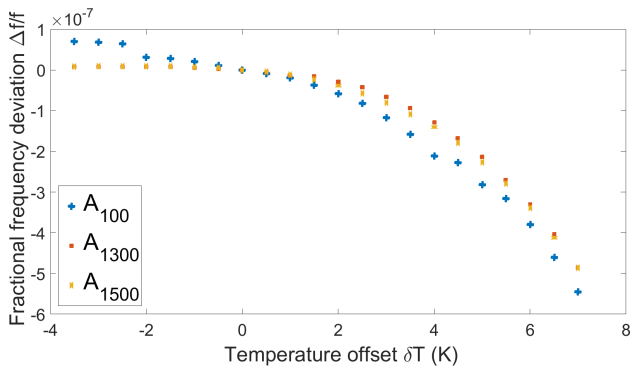}
	\captionof{figure}{Resonance fractional frequency shift $\frac{f(T)-f(T_{ref})}{f(T_{ref})}$ for a X-cut as a function of temperature offset $\delta T = T-T_{ref}$, where $T_{ref} = 8 {\rm K}$.}
	\label{fig:F_T_curves}
\end{figure}

As an example, Fig.~\ref{fig:F_T_curves} shows experimental results of the fractional frequency deviation versus temperature for a X-cut operating around $T_{ref}=8K$. The geometrical properties of this device are $h = 600 \mu m$ and $R_0 = 250$ mm. It can be observed that all the A-mode  overtones exhibit similar behaviors.The temperature coefficients of the frequency are extracted from such experimental data (see Tab.~\ref{tab:table1}).
As a final result, Tab.~\ref{tab:table2} gives the thermal coefficients $T^{(k)}C_{ij}$ of the elastic coefficients $C_{ij}$ of quartz, up to the $3^{rd}$ order, determined according to the Eulerian description.



\begin{table} [h!]
\caption{\label{tab:table2}$1^{st}$, $2^{nd}$, and $3^{rd}$ order temperature coefficients of the elastic constants of crystalline quartz within 4 {\rm K} - 15 {\rm K}, with $T_{ref} = 8 {\rm K}$.}
\begin{ruledtabular}
\begin{tabular}{ccccc}
$C_{ij}$ & $T^{(1)}C_{ijkl}, 10^{-7}$ & $T^{(2)}C_{ijkl} 10^{-8}$ & $T^{(3)}C_{ijkl}, 10^{-9}$\\
\hline
$C_{11}$ & $-0.235$ & $1.28$ & $-3.25$ \\
$C_{12}$\footnotemark[1] & $-6.39$ & $-10.56$ & $-4.22$ \\
$C_{13}$ & $-7.18$ & $-7.42$ & $-0.877$ \\
$C_{14}$ & $-1.35$ & $-1.31$ & $-7.93$ \\
$C_{33}$ & $-4.62$ & $-5.65$ & $-7.62$ \\
$C_{44}$ & $-3.17$ & $-4.37$ & $-11.6$ \\
$C_{66}$ & $3.08$ & $5.92$ & $0.486$ \\
\end{tabular}
\end{ruledtabular}
\footnotetext[1]{Calculated coefficient}
\end{table}

\section{Conclusion}

Various cut-angles of synthetic quartz have been measured at cryogenic temperature in order to determine the temperature coefficients of its elastic coefficients. The Eulerian formalism have been used to provide these $T^{(k)}C_{ij}$ up to the $3^{rd}$ order. Uncertainties on the elastic coefficient values are not specified  because most of the uncertainties linked to the many steps of measurements and calculation are difficult to bring out. Actually, in addition to the cut angle uncertainty, one of the higher uncertainties is the result of the Tiersten model which is a theoretical approach modelling plano-convex resonators. However, it can be stated that the uncertainty on the temperature sensitivity of the elastic coefficient $C_{13}$ is higher than that of other coefficients because of its very slight involvement in calculations.

\section*{Acknowledgements}
	
Thanks to Philippe Abb\'{e} and Xavier Vacheret for their help for making resonators and experiments. This work is supported by Conseil R\'{e}gional de Franche-Comt\'{e}. J.B. is thankful to Minist\`{e}re de l'Enseignement Sup\'{e}rieur et de la Recherche, France, for his grant.

\bibliography{biblio}

\end{document}